\documentclass{article}
\usepackage{spconf,amsmath,graphicx}
\usepackage{caption}
\usepackage{footmisc}
\usepackage{graphicx}
\usepackage{float} 
\usepackage{subfigure}
\usepackage{subcaption}
\usepackage{subfig}
\usepackage{amssymb,amsmath,bm}
\usepackage{enumitem}
\usepackage{multirow} 
\usepackage{makecell}
\usepackage{bbding}
\usepackage{pifont}
\usepackage{wasysym}
\usepackage{twemojis}

\title{Pitch-and-Spectrum-Aware Singing Quality Assessment with Bias Correction and Model Fusion}
\name{Yu-Fei Shi, Yang Ai\sthanks{ Corresponding author. This work was funded by the National Nature Science Foundation of China under Grant 62301521, the Anhui Provincial Natural Science Foundation under Grant 2308085QF200, and the Fundamental Research Funds for the Central Universities under Grant WK2100000033.}, Ye-Xin Lu, Hui-Peng Du, Zhen-Hua Ling}

\address{National Engineering Research Center of Speech and Language Information Processing, \\University of Science and Technology of China, Hefei, P. R. China \\{\small \tt \ \{zkddsr2023, yxlu0102, redmist\}@mail.ustc.edu.cn, \{yangai, zhling\}@ustc.edu.cn}}

\begin{document}
\ninept
\maketitle

\begin{abstract}
We participated in track 2 of the VoiceMOS Challenge 2024, which aimed to predict the mean opinion score (MOS) of singing samples. 
Our submission secured the first place among all participating teams, excluding the official baseline.
In this paper, we further improve our submission and propose a novel Pitch-and-Spectrum-aware Singing Quality Assessment (PS-SQA) method. 
The PS-SQA is designed based on the self-supervised-learning (SSL) MOS predictor, incorporating singing pitch and spectral information, which are extracted using pitch histogram and non-quantized neural codec, respectively. 
Additionally, the PS-SQA introduces a bias correction strategy to address prediction biases caused by low-resource training samples, and employs model fusion technology to further enhance prediction accuracy. 
Experimental results confirm that our proposed PS-SQA significantly outperforms all competing systems across all system-level metrics, confirming its strong sing quality assessment capabilities.

\end{abstract}
\section{Introduction}
\label{sec:intro}

With the rapid development of singing voice synthesis (SVS) and singing voice conversion (SVC) systems, there is an urgent need for technology that can automatically assess the quality of generated singing voice, instead of traditional subjective listener scoring methods which are time-consuming and inefficient. 
However, in past research on singing quality assessment, most studies focus on the quality assessment of real recorded human singing voices \cite{huang2020spectral,li2021training,sun2023tg}. 
To our knowledge, the quality assessment of generated singing voices has not yet been thoroughly investigated. 
Nonetheless, methods for speech quality assessment can serve as references and be applied to singing quality assessment.
The mean opinion score (MOS) is the gold standard in the fields of speech synthesis and voice conversion \cite{black2005blizzard}, representing the average five-point rating given by humans to generated speech. 
Early MOS prediction models employed bidirectional long short-term memory recurrent neural network (BiLSTM-RNN) or convolutional neural network (CNN) to predict MOS score from input speech waveforms or amplitude spectra \cite{patton2016automos,fu2018quality,lo2019mosnet}. 
Recently, with the development of self-supervised learning (SSL) methods, fine-tuning SSL models and adapting them for speech MOS prediction has become one of the state-of-the-art approaches.
MOS prediction is also applicable for singing quality assessment, but directly transferring methods from speech MOS prediction is clearly inappropriate, as they may not align with the characteristics of singing voices.

The VoiceMOS Challenge 2024, launched this year, aims to encourage participants to build systems that predict MOS of singing voices generated by SVS and SVC systems in track 2. 
The challenge provides a platform for innovators to develop and test their models against a standardized dataset, advancing the progress of singing quality assessment. 
The organizers provide a singing quality evaluation dataset, SingMOS, along with a baseline. 
Participants are required to build systems to predict the MOS for the singing voices in the evaluation set, and the organizers rank the participating systems using certain metrics. 
Our submitted system secured first place among all participating teams (excluding the official baseline). 
Nevertheless, there is still significant room for improvement in the accuracy of singing MOS prediction.

To address the existing issues in singing quality assessment, this paper further improves our competition system submitted to track 2 of VoiceMOS Challenge 2024 and proposes a novel Pitch-and-Spectrum-aware Singing Quality Assessment (PS-SQA) method. 
To address the characteristics of singing voices, the PS-SQA innovatively introduces pitch-aware SSL-based MOS predictors and spectrum-aware SSL-based MOS predictors. 
They are built on the plain SSL-based MOS predictor by introducing pitch histograms and spectral-level acoustic features encoded by a non-quantized APCodec \cite{ai2024apcodec}, respectively. 
These methods attempt to inject key singing information such as musical melody into the predictor, providing a new approach suitable for evaluating synthesized singing. 
These methods enable the predictor to have a more nuanced understanding of the melodic content of the singing, which is crucial for accurate quality assessment. 
Additionally, we noticed that the official SingMOS dataset has the issue of imbalanced training sample distribution. 
To address this, the PS-SQA introduces a bias correction branch, integrating into aforementioned predictors to mitigate the effects of such imbalances. 
Finally, PS-SQA also employs a model fusion strategy, comprehensively considering the results of multiple MOS predictors to provide a more thorough and accurate singing MOS score. 
Experimental results confirm that our proposed PS-SQA significantly improves our submission system and clearly outperforms all competing systems in track 2 of VoiceMOS Challenge 2024 in terms of the system-level spearman rank correlation coefficient (SRCC) used for ranking.


This paper is organized as follows:
In Section \ref{sec: Related Works}, We provide a brief review of the related work involved in PS-SQA, encompassing SSL-based MOS predictors, pitch histograms, and advanced neural audio codecs. 
In Section \ref{sec: Proposed Methods}, we give details of the construction and workflow of the various components that make up PS-SQA. 
In Section \ref{sec: Experiments}, we present our experimental results.
Finally, we give conclusions in Section \ref{sec: Conclusion}.

\section{Related Work}
\label{sec: Related Works}
\subsection{SSL-based MOS Predictor}
\label{ssl}
\begin{figure}[t]
    \centering
    \includegraphics[scale=0.4]{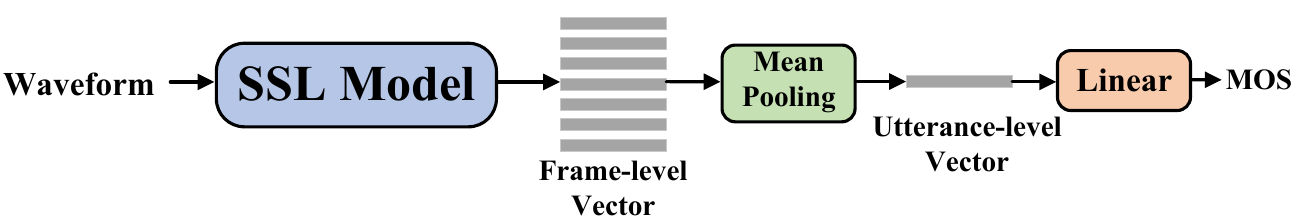}
    \caption{A block diagram of a plain SSL-based MOS predictor.}
    \label{fig:ssl}
\end{figure}

Recently, SSL models trained with a large amount of unlabeled data using self-supervised learning have been applied to MOS prediction, achieving impressive results \cite{cooper2022generalization}.
In both the VoiceMOS Challenge 2022 and 2023, top-ranking teams employed fine-tuning on SSL models to achieve perfect predictive accuracy \cite{huang2022voicemos,cooper2023voicemos}. 
Therefore, in the VoiceMOS Challenge 2024, we also fine-tuned the official baseline SSL-based MOS predictor to develop our system. 
The process of predicting MOS using SSL-based predictor is illustrated in Figure \ref{fig:ssl}. 
The waveform is processed through a pre-trained SSL model to produce a frame-level feature vector, which is then averaged using mean-pooling to obtain an utterance-level one. 
Finally, a linear layer reduces the feature dimensionality to 1 to derive the corresponding MOS score. 
Assuming $\hat{y}$ is the predicted MOS and $y$ is its corresponding label, the loss function is defined as the L1 error between $\hat{y}$ and $y$, i.e.,
\begin{equation} 
\mathcal L_{ssl} = \Vert\hat{y} - y\Vert_1.
\tag{1}\label{eq1}\end{equation}


\subsection{Pitch Histogram}
\label{pitch-histogram}

\begin{figure}[t]
\setlength{\belowcaptionskip}{-0.5cm}
\begin{center}
\subfigure[]{\includegraphics[width=1.6in,height=1in]{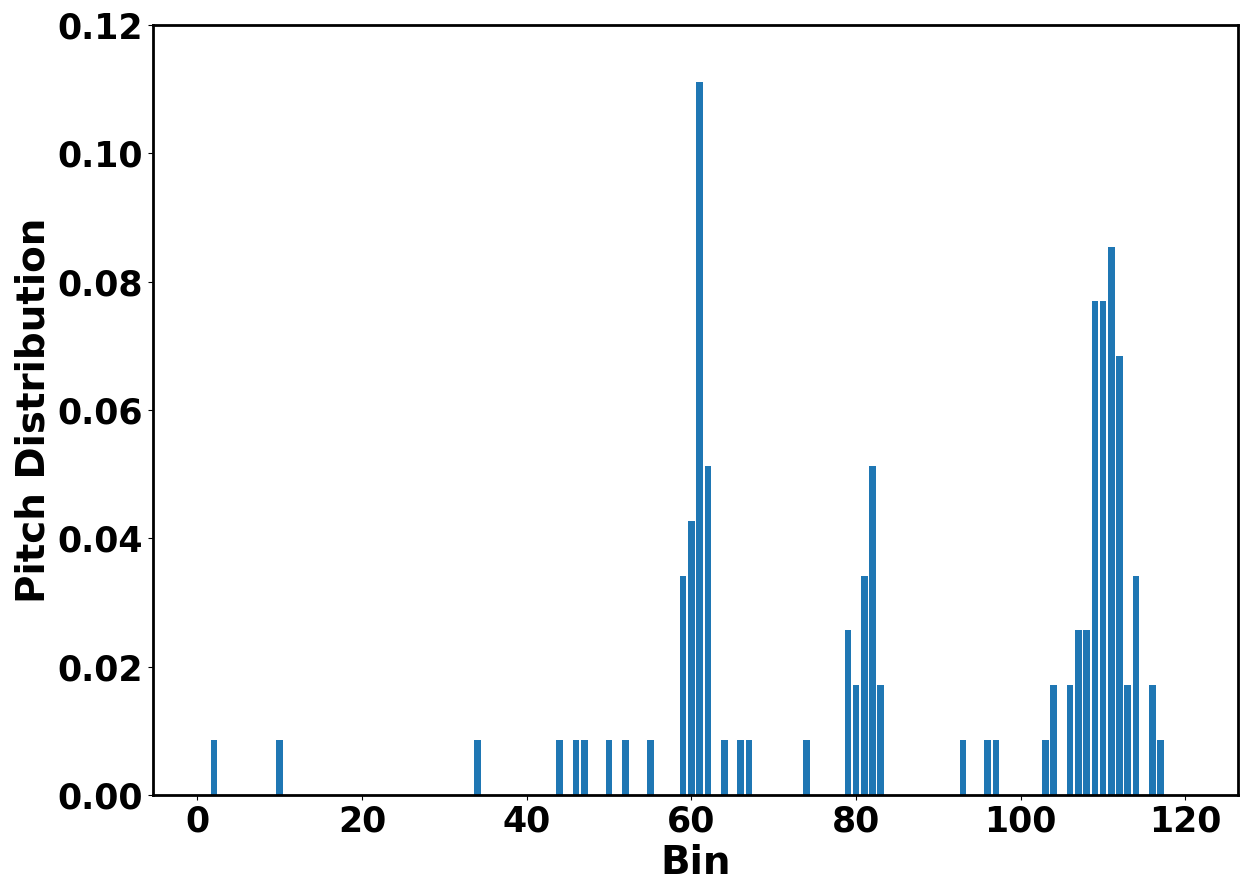}}
\subfigure[]{\includegraphics[width=1.6in,height=1in]{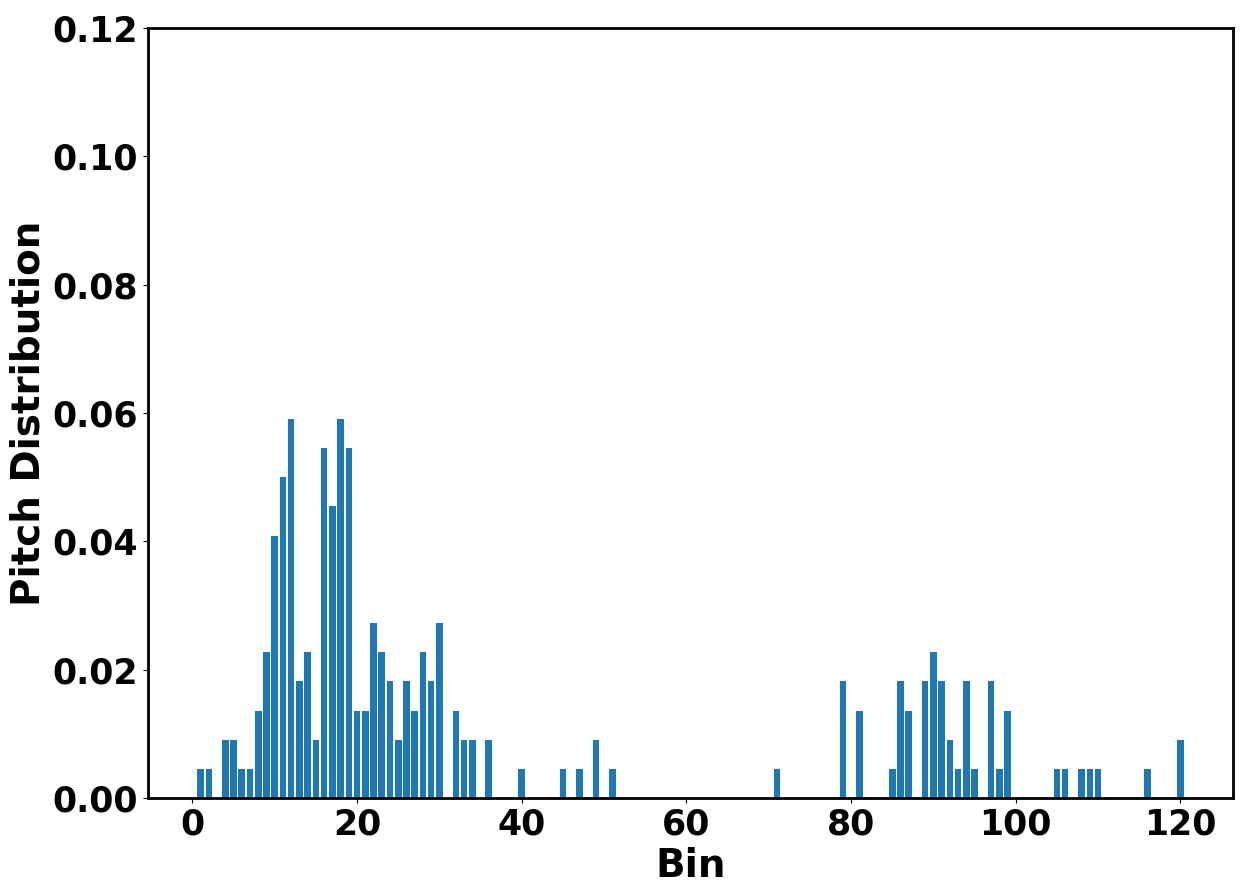}}
\caption{The pitch histograms of (a) a good singing voice with MOS of 5.0 and (b) a poor singing voice with MOS of 2.4.}
\label{histograms}
\end{center}
\end{figure}

Pitch is a critical measure in evaluating singing quality. 
Unlike direct use of detected fundamental frequency values in speech processing, assessing musical quality often involves transforming pitch values. 
Since adjacent notes in sheet music have consistent pitch ratios, folding identified pitch values over an octave can accurately reconstruct the melody of a singing voice. 
Specifically, in the MIDI scale, a complete octave contains 12 semitones, with the pitch ratio between adjacent semitones being $2^{1/12}$ and the pitch frequency ratio between each octave is $2$. 
Additionally, we further divide the interval between adjacent semitones into 100 cents and the pitch frequency ratio between adjacent cents is $2^{1/1200}$. 
On this basis, we convert the aforementioned pitch frequency from the Hz scale ($f_{Hz}$) to the cent scale ($f_{cent}$) using the following formula:
\begin{equation}
f_{cent} = 1200\times \log_2 \frac {f_{Hz}}{440},
\tag{2}\label{eq2}
\end{equation}
where 440 Hz (pitch-standard musical note A4) is considered as the base frequency.

In our implementation, we first use \textit{PyWORLD} to extract the pitch in Hz (i.e., $f_{Hz}$) of different frames of singing. Then, using Equation \ref{eq2}, we get the converted pitch $f_{cent}$ in the unit of cents. This allows us to obtain a pitch sequence that better reflects the characteristics of the singing voice. 
Following prior work \cite{huang2020spectral,tzanetakis2003pitch,gupta2017perceptual}, we treat 10 cents as the smallest counting unit (i.e., 1 bin). 
Therefore, there are a total of 12 semitones × 10 bins = 120 bins in an octave. 
Then, we further convert the pitch value $f_{cent}$ to a compressed continuous value $I(f_{cent})\in [0,120)$, i.e., 
\begin{equation}
I(f_{cent}) = \frac{f_{cent}}{10} \bmod 120,
\tag{3}\label{eq3}
\end{equation}
where $\bmod$ denotes the modulo operation. 

\graphicspath{ {plot/} }
\begin{figure}[t]
\setlength{\belowcaptionskip}{-0.5cm}
\centering

\subfigure[ Compressed-pitch-aware SSL-based MOS predictor]{
		\includegraphics[scale=0.31]{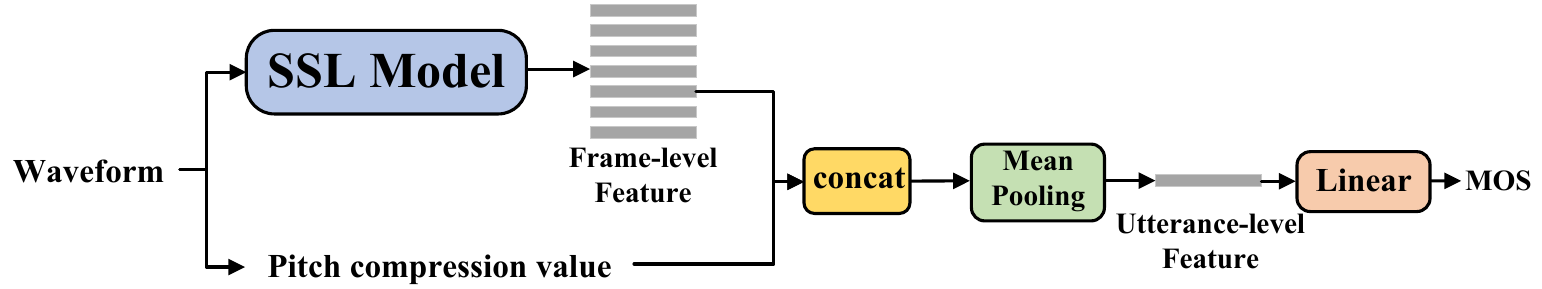}}\\
\subfigure[ Pitch-histogram-aware SSL-based MOS predictor]{
		\includegraphics[scale=0.274]{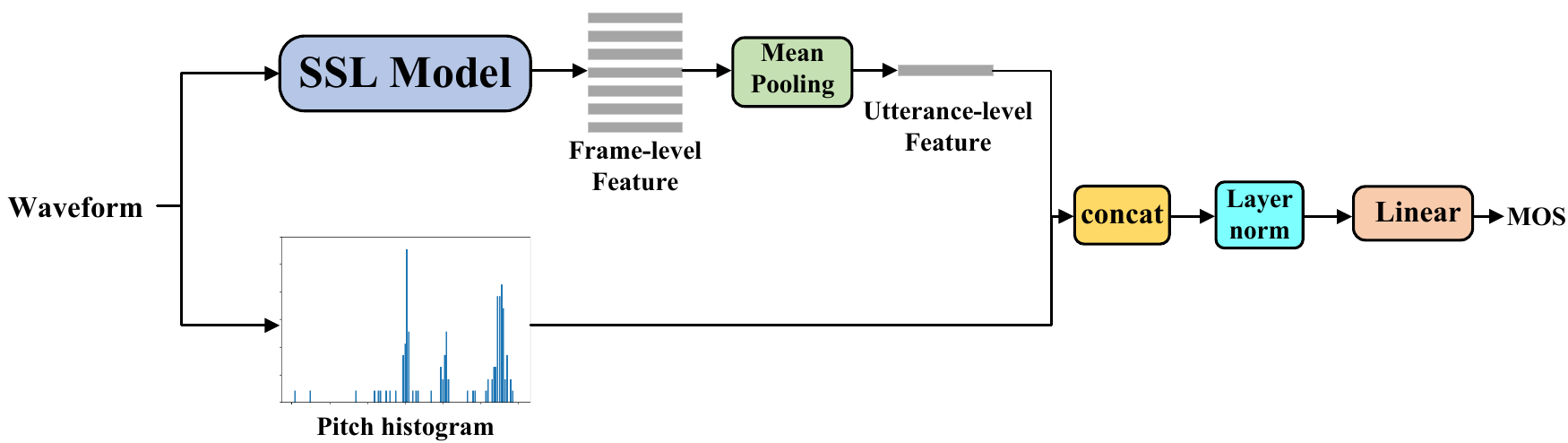}}
\subfigure[ Spectrum-aware SSL-based MOS predictor]{
		\includegraphics[scale=0.306]{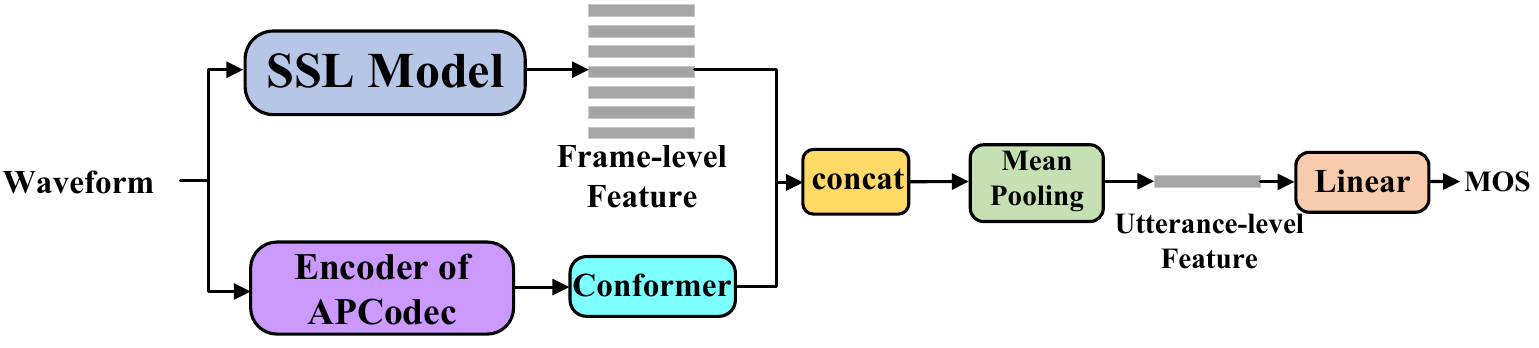}}
\caption{ Architectures of the pitch-aware and spectrum-aware SSL-based MOS predictors.}
\label{PA/SA-SSL-MOS}
\end{figure}
In this way, pitch values from different frames of a singing voice can all be mapped into one octave. 
Subsequently, the pitch histogram $\bm{P}=\{P_1,\dots,P_j,\dots,P_{120} \}$ of a sample is obtained by calculating the ratio of the frame-level pitch count in each bin to the total number of frames, i.e.,
\begin{equation}
P_j =\dfrac{1}{N} \sum_{n=1}^{N} m_j^n,
\tag{4}\label{eq4}
\end{equation}
where
\begin{align}
\label{eq5}
\tag{5}
m_j^n=\left\{\begin{array}{ll}1,& j-1 \leq I(f_{cent}^n) < j\\ 0,&\text{otherwise}\end{array}\right..
\end{align}
Here, $f_{cent}^n$ represents the pitch value in cent of the $n$-th frame and $N$ is the total number of frames of a sing sample.


The melody of a song is typically composed of a set of pitch values that frequently appear in the vocals. 
In previous work \cite{huang2020spectral,gupta2018automatic}, it was found that for different singers performing the same song, the pitch histograms of good singers exhibit sharper peaks. 
This indicates that the notes of the song are consistently hit. 
In contrast, the pitch histograms of poor singers do not show such prominent peaks because they fail to consistently hit the dominant notes, resulting in being out of tune. 
Figure \ref{histograms} shows pitch histograms extracted from two samples provided in track 2 of VoiceMOS Challenge 2024, where it can be seen that the histogram of the high-MOS singing voice has more pronounced peaks compared to that of the low-MOS singing voice. 
This is why we chose the pitch histogram to assist in sing quality assessment in PS-SQA. 

\subsection{Neural Audio Codec}
\label{apcodec}
Neural audio codec is a crucial signal processing technology that compresses audio signals into discrete codes and then reconstructs the original audio from these codes by neural networks. 
It can also be considered a feature extractor, capable of effectively extracting intermediate representations that contain acoustic information from the audio, which can be used as input for MOS prediction models. 
Early neural audio codecs, such as SoundStream \cite{zeghidour2021soundstream} and Encodec \cite{defossez2023high}, directly encode the time-domain audio waveforms. 
Recently, Ai \MakeLowercase{\textit{et al.}} proposes APCodec \cite{ai2024apcodec}, a parametric neural audio codec that uses the audio amplitude and phase spectra as coding objects. 
Therefore, through explicit spectral modeling, the intermediate features encoded by APCodec contain richer spectral-level acoustic information. 
This is related to several amplitude-spectrum-based MOS prediction methods \cite{lo2019mosnet,zezario2022deep}, but it leverages the missing phase spectrum information in these methods, making it more suitable for MOS prediction. 
This is why we chose APCodec to extract spectral features to assist in singing quality assessment in PS-SQA.

\begin{figure}
\setlength{\belowcaptionskip}{-0.5cm}
\begin{center}
\subfigure[]{\includegraphics[width=0.48\linewidth]{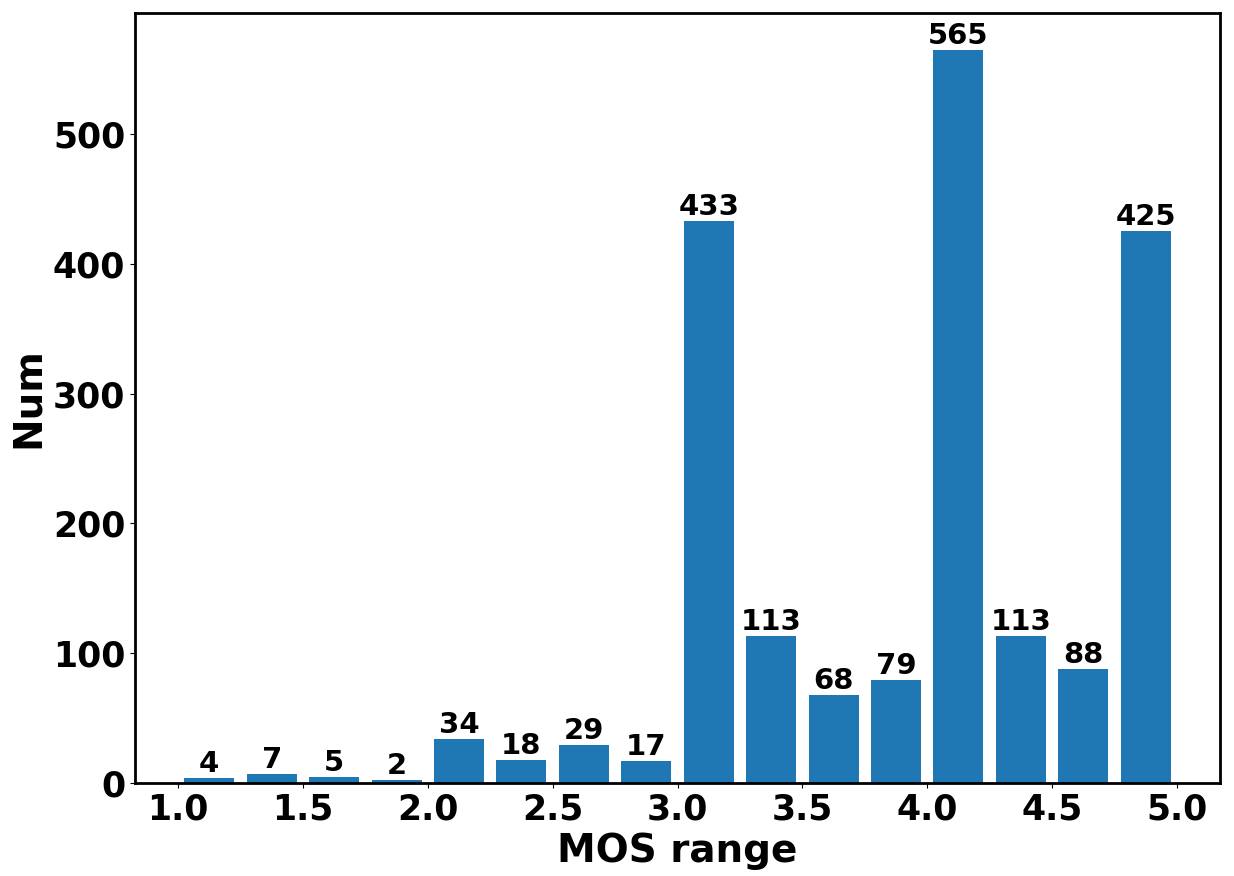}}
\subfigure[]{\includegraphics[width=0.48\linewidth]{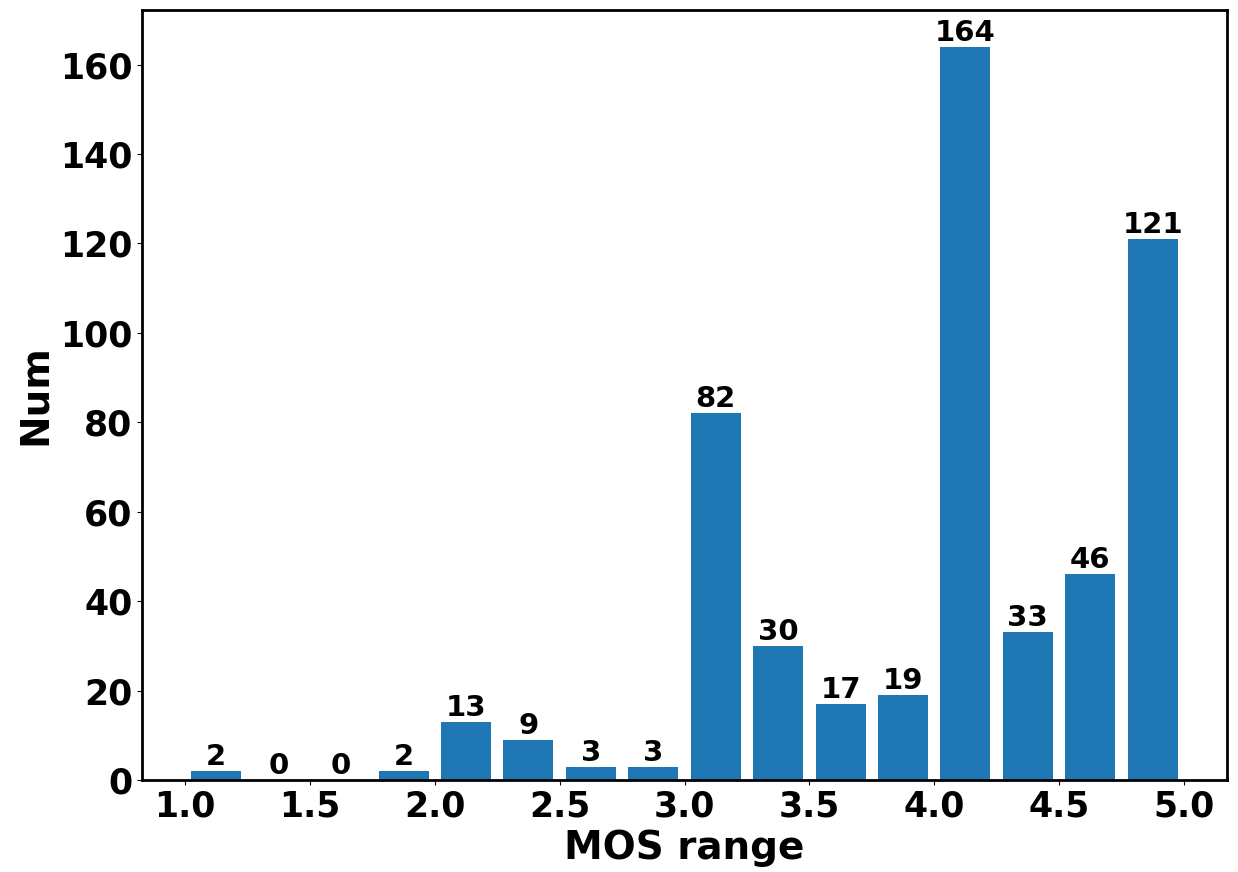}}
\caption{
Histograms of the number of samples in different MOS intervals for (a) training set and (b) validation set of SingMOS dataset.}
\label{distribute of dataset}
\end{center}
\end{figure}

\section{PROPOSED METHOD}
\label{sec: Proposed Methods} 

The core of PS-SQA lies in introducing pitch-aware SSL-based MOS predictors and spectrum-aware SSL-based MOS predictors based on the plain SSL-based MOS prediction framework, tailored for quality assessment that suits the characteristics of singing voices. 
Additionally, to overcome the issues caused by imbalanced training data, PS-SQA innovatively introduces a bias correction strategy. 
Finally, PS-SQA also employs model fusion techniques, aggregating the results of multiple top-ranking MOS predictors to output a comprehensive MOS score, further enhancing the predictive accuracy of PS-SQA. 
All predictors are trained using the loss function defined in Equation \ref{eq1}.


\vspace{-1mm}
\subsection{Pitch-aware SSL-based MOS predictor}
\vspace{-1mm}

As mentioned in Section \ref{ssl}, the SSL models are widely used in speech MOS prediction. 
The SSL models extract semantic information from speech waveforms for MOS prediction, which is clearly not suitable for singing quality assessment. 
Considering the strong relevance of pitch to singing quality discussed in Section \ref{pitch-histogram}, we propose to integrate pitch information into the plain SSL-based MOS predictor to enhance MOS prediction accuracy for singing quality assessment. 
By explicitly providing pitch-related information to the network, this approach also releases network degrees of freedom to focus on learning non-pitch related properties.

Therefore, the PS-SQA designs pitch-aware SSL-based MOS predictors by incorporating pitch information into the plain SSL-based MOS prediction framework in a specific manner. 
Furthermore, by adopting different forms of pitch information, we attempt to construct a compressed-pitch-aware SSL-based MOS predictor and a pitch-histogram-aware SSL-based MOS predictor, respectively. 
\begin{itemize}[nosep, leftmargin=*]
\item {}{\textbf{Compressed-pitch-aware SSL-based MOS predictor:}} As shown in Figure \ref{PA/SA-SSL-MOS}(a), the compressed-pitch-aware SSL-based MOS predictor integrates the compressed value-constrained pitch sequence $\bm{f}_c=\left[ I(f_{cent}^1),\dots,I(f_{cent}^N)\right]^\top$ into a plain SSL-based MOS prediction framework. 
Specifically, the SSL model extracts frame-level features from the waveform using the same frame shift as \textit{PyWORLD} does when extracting the pitch from the waveform. 
This ensures that the two extracted sequences have the same temporal resolution. 
Then, we concatenate the SSL-model-processed frame-level features with $\bm{f}_c$ along the dimension axis. 
After pooling along the time axis to obtain the utterance-level feature, we pass it through a linear layer to output the 1-dimensional MOS score.

\item {}{\textbf{Pitch-histogram-aware SSL-based MOS predictor:}} As shown in Figure \ref{PA/SA-SSL-MOS}(b), the pitch-histogram-aware SSL-based MOS predictor uses a pitch histogram as a conditioning vector for plain SSL-based MOS prediction framework. 
Since the pitch histogram statistically represents the pitch distribution of an utterance, referring to \cite{gupta2020automatic}, it is concatenated with the utterance-level features obtained by pooling the frame-level features outputted by the SSL model. 
This is different from the aforementioned compressed-pitch-aware SSL-based MOS predictor. 
Then we balance these two types of features using a layer normalization and finally output the 1-dimensional MOS score through a linear layer.

\end{itemize}

\vspace{-1mm}
\subsection{Spectrum-aware SSL-based MOS predictor}
\vspace{-1mm}

As mentioned in Section \ref{apcodec}, neural audio codecs differ from SSL models in that they can extract acoustic features from audio waveforms, and acoustic information is crucial for singing quality assessment. 
We use APCodec \cite{ai2024apcodec} to extract spectral-level acoustic features because it uses the audio amplitude and phase spectra as coding objects. 
In our preliminary experiments, we found that the quantization process affects the quality of the extracted features.
While quantization is unavoidable for audio compression, it is not necessary for quality assessment. 
Therefore, we discard the quantizer and construct a non-quantized APCodec, consisting only of an encoder and a decoder. 
As depicted in Figure \ref{PA/SA-SSL-MOS}(c), for spectrum-aware SSL-based MOS predictor, a well-trained APCodec encoder extracts spectral-level acoustic features from singing voice waveforms. 
The frame shift used in this operation is the same as that used in the SSL model. 
Following this, a Conformer \cite{gulati2020conformer} is leveraged to capture a global view of the spectral representations from the input acoustic features. 
Similar to compressed-pitch-aware SSL-based MOS predictor, the output of the Conformer is concatenated with the output of the SSL model along the dimension axis, and then passed through a pooling layer and a linear layer to produce the 1-dimensional MOS score.


\vspace{-2.5mm}
\subsection{Bias Correction Branch}
\label{bias_net}
\vspace{-1mm}
We propose a bias correction branch to replace the linear layer that outputs the MOS score in SSL-based MOS predictors, aiming at addressing the issue of imbalanced training sample distribution. 
As shown in Figure \ref{distribute of dataset}, we conducted a quantitative analysis of the SingMOS dataset provided by track 2 of VoiceMOS Challenge 2024 based on the score intervals of the labels. 
It can be observed that there are few samples in the low MOS intervals in both the training and validation sets, with the majority of training samples concentrated in the MOS interval above 3.0. 
Some other datasets \cite{huang2022voicemos} have MOS scores concentrated in the middle range, lacking high-MOS samples.
We hypothesize that this imbalance in the training sample distribution may cause the trained model to have weaker predictive capabilities for samples with labels in certain MOS ranges compared to others, resulting in significant deviations between the predicted MOS and the actual labels.

\begin{figure}[t]
\setlength{\belowcaptionskip}{-0.5cm}
    \centering
    \includegraphics[scale=0.45]{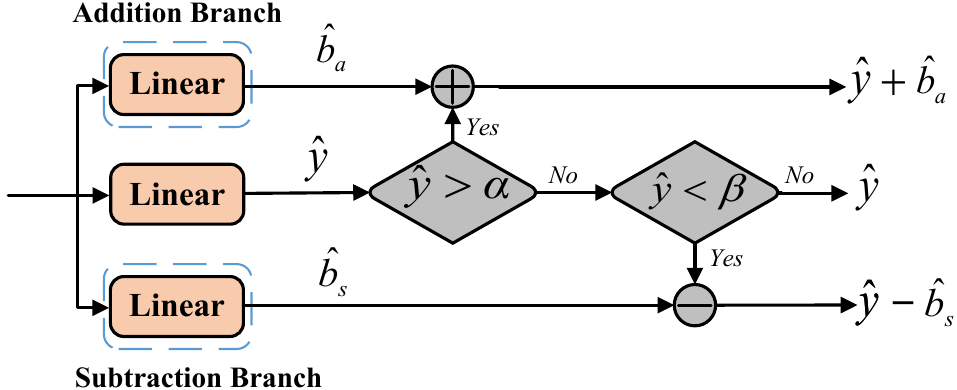}
    \caption{The structure of the bias-correction branch.}
    \label{fig:bias-ssl}
\end{figure}

To address this, we design a bias correction branch composed of three parallel linear layers that can theoretically be applied to any MOS prediction model by attaching it to the model's output. 
As shown in Figure \ref{fig:bias-ssl}, the main improvement of the bias correction branch, compared to the commonly used linear layer at the output end of SSL-based MOS predictors, is the introduction of addition branch and subtraction branch. 
One branch is responsible for correcting the predicted scores of high MOS samples, while the other branch corrects the predicted scores of low MOS samples. 
Assume that the middle linear layer outputs the original MOS score $\hat{y}$, while addition branch and subtraction branch output bias values $\hat{b}_a$ and $\hat{b}_s$, respectively. 
Given MOS score thresholds $\alpha$ and $\beta$ ($1<\beta<\alpha<5$), the final output MOS score $\hat{y}_{bc}$ of the bias correction branch is defined as 
\begin{align}
\label{eq6}
\tag{6}
\hat{y}_{bc}=\left\{\begin{array}{ll}\hat{y}+\hat{b}_a,& \hat{y}>\alpha\\ \hat{y},&\beta<\hat{y}<\alpha \\ \hat{y}-\hat{b}_s, & \hat{y}<\beta\end{array}\right..
\end{align} 
This approach can prevent the MOS predictor from being confined to the middle MOS score range, allowing it to break through the interval boundaries and improve its assessment capabilities for both high and low MOS score samples. 
In the actual training process, after finishing training the original SSL-based MOS predictor, addition and subtraction branches are introduced at the model's output, and only these two branches are trained while keeping the other model parameters fixed.



\graphicspath{ {plot/} }
\vspace{-1mm}
\subsection{Model Fusion}
\label{fusion} 
\vspace{-1mm}
According to previous studies \cite{kunevsova2023ensemble, yang2022fusion,stan22_interspeech,qi2023ssl}, it is evident that different types of MOS predictors can capture various information present in the training set. 
Integrating these models can effectively leverage these diverse perspectives to enhance the predictive capabilities of the quality assessment system. 
Therefore, we employ multiple pre-trained SSL models from $\textit{fairseq}$ \cite{ott2019fairseq} as our SSL models in plain, pitch-aware and spectrum-aware SSL-based MOS predictors. 
Then, we rank all MOS predictors according to preset rules and select the top predictors. 
After introducing the bias correction branches to each selected predictor, we perform the model fusion operation.
Specifically, we concatenate the outputs of selected predictors into a single fusion vector as input features, then a linear layer acts as the combiner to output the final fused MOS score.

\section{Experiments}
\label{sec: Experiments}
\subsection{Dataset and Evaluation Metrics}
During the training phase, the track 2 of the VoiceMOS Challenge 2024 released a dataset named SingMOS\cite{tang2024singmos}. 
This dataset comprises Mandarin and Japanese samples obtained from SVS systems, SVC systems, analysis-synthesis operation of neural vocoders, and natural singing voice recordings. 
In total, the dataset includes 3189 samples, with 2000 samples allocated to the training set, 544 samples to the validation set, and 645 samples to the evaluation set. 
Notably, the validation and evaluation sets include unseen samples.

The evaluation metrics encompass mean squared error (MSE), linear correlation coefficient (LCC), SRCC, and kendall tau rank correlation (KTAU) at both the utterance and system levels \cite{lo2019mosnet}. 
MSE quantifies the disparity between predicted and ground truth MOS scores, whereas the other metrics assess correlation. 
According to the rules of the track 2 of VoiceMOS Challenge 2024, the system-level SRCC is used as the ranking criterion.

\begin{table*}[t]
 \setlength{\tabcolsep}{3pt}
 \caption{
Experimental results of plain SSL-based MOS predictor (SSL-MOS), compressed-pitch-aware SSL-based MOS predictor (CP-SSL-MOS), pitch-histogram-aware SSL-based MOS predictor (PH-SSL-MOS) and spectrum-aware SSL-based MOS predictor (S-SSL-MOS) when adopting different SSL models evaluated on the validation set of SingMOS. The \textbf{bold} and \underline{underline} numbers indicate the optimal and sub-optimal results, respectively. 
 }
 \label{table:main experiment}
 \centering
 \begin{tabular}{l c c c c c |c c c c } 
  \hline
  \multirow{2}{*}{\textbf{Predictor}}&\multirow{2}{*}{\textbf{SSL Model}}&\multicolumn{4}{c|}{\textbf{Utterance-level Metrics}} &\multicolumn{4}{c}{\textbf{System-level Metrics}}\\
  \cline{3-10}
   & & MSE $\downarrow$ & LCC $\uparrow$ & SRCC $\uparrow$ & KTAU $\uparrow$ & MSE $\downarrow$ & LCC $\uparrow$ & SRCC $\uparrow$ & KTAU $\uparrow$\\  
  \hline
  
   SSL-MOS& \multirow{4}{*}{\makecell[c]{Wav2Vec2.0 Base}} & 0.470&0.652&\textbf{0.647}&\textbf{0.487}&\underline{0.126}&0.945&0.928&0.775\\

  CP-SSL-MOS&  &\underline{0.370}&\underline{0.668}&\underline{0.635}&\underline{0.475}&\textbf{0.036}&\textbf{0.952}&\textbf{0.939}&\textbf{0.813} \\

  PH-SSL-MOS&  &\textbf{0.344}&\textbf{0.669}&0.629&0.474&0.241&\underline{0.951}&\textbf{0.939}&\underline{0.806}\\

  S-SSL-MOS&  &0.477&0.660&0.623&0.465&0.159&0.908&\underline{0.931}&0.783\\

  \hline  
  
  SSL-MOS & \multirow{4}{*}{\makecell[c]{Wav2Vec2.0 Large}} & \underline{0.458}&\underline{0.629}&\underline{0.604}&\underline{0.448}&\underline{0.117}&\textbf{0.939}&\underline{0.922}&\underline{0.768}\\

  CP-SSL-MOS&  &0.654&0.261&0.281&0.199&0.325&0.603&0.606&0.445\\
  
  PH-SSL-MOS&  &\textbf{0.320}&\textbf{0.689}&\textbf{0.651}&\textbf{0.489}&\textbf{0.035}&\underline{0.935}&\textbf{0.933}&\textbf{0.787} \\

  S-SSL-MOS&  &0.543&0.501&0.469&0.334&0.202&0.780&0.749&0.571\\

  \hline

SSL-MOS & \multirow{4}{*}{\makecell[c]{HuBERT Base}} & 0.406&0.642&0.619&0.459&0.077&0.925&0.910&0.749\\

  CP-SSL-MOS&  &\underline{0.383}&0.612&0.611&0.450&\textbf{0.036}&0.932&0.921&0.775\\
  
  PH-SSL-MOS&  &\textbf{0.338}&\textbf{0.685}&\textbf{0.653}&\textbf{0.489}&\underline{0.051}&\textbf{0.949}&\textbf{0.935}&\textbf{0.806}\\

  S-SSL-MOS&  &0.966&\underline{0.656}&\underline{0.627}&\underline{0.465}&0.620&\underline{0.945}&\underline{0.931}&\underline{0.798}\\
  
  \hline

SSL-MOS & \multirow{4}{*}{\makecell[c]{HuBERT Large}} &0.485&0.533&0.524&0.378&0.093&0.823&0.855&0.688\\

  CP-SSL-MOS&  &\underline{0.434}&0.553&0.539&0.397&\underline{0.092}&0.874&0.876&0.715\\
  
  PH-SSL-MOS&  &\textbf{0.399}&\underline{0.588}&\underline{0.573}&\underline{0.420}&\textbf{0.059}&\textbf{0.904}&\underline{0.894}&\underline{0.726}\\

  S-SSL-MOS&  &1.387&\textbf{0.616}&\textbf{0.586}&\textbf{0.429}&1.026&\underline{0.886}&\textbf{0.896}&\textbf{0.730}\\
  
  \hline

SSL-MOS & \multirow{4}{*}{\makecell[c]{HuBERT Extra Large}} &0.443&0.543&0.539&0.395&0.064&0.876&0.887&0.715\\

  CP-SSL-MOS&  &\underline{0.421}&0.602&0.599&0.440&\underline{0.048}&0.909&\underline{0.918}&0.765\\
  
  PH-SSL-MOS&  &\textbf{0.359}&\textbf{0.651}&\textbf{0.641}&\textbf{0.478}&\textbf{0.043}&\underline{0.919}&\textbf{0.923}&\underline{0.772}\\

  S-SSL-MOS&  &1.873&\underline{0.635}&\underline{0.600}&\underline{0.442}&1.274&\textbf{0.936}&\textbf{0.923}&\textbf{0.779}\\
  
  \hline

 \end{tabular}
\end{table*}

\begin{table*}[t]
 \setlength{\tabcolsep}{3pt}
 \caption{  
Experimental results of the fused model of PS-SQA and included predictors without or with bias correction branch evaluated on the validation set of SingMOS. The \textbf{bold} number indicates the optimal results. 
 }
 \label{table:sub experiment}
 \centering
 \begin{tabular}{l c c c c c c |c c c c } 
  \hline
  \multirow{2}{*}{\textbf{Fusion Model \& Predictor}}&\multirow{2}{*}{\textbf{SSL Model}}&\multirow{2}{*}{\textbf{Bias Correction}}&\multicolumn{4}{c|}{\textbf{Utterance-level Metrics}} &\multicolumn{4}{c}{\textbf{System-level Metrics}}\\
  \cline{4-11}
   & && MSE $\downarrow$ & LCC $\uparrow$ & SRCC $\uparrow$ & KTAU $\uparrow$ & MSE $\downarrow$ & LCC $\uparrow$ & SRCC $\uparrow$ & KTAU $\uparrow$\\  
  \hline
  
   \textbf{PS-SQA}& \- & \twemoji{multiply}&\textbf{0.302}&\textbf{0.710}&\textbf{0.674}&\textbf{0.507}&\textbf{0.031}&0.946&\textbf{0.952}&\textbf{0.840}\\

  \quad \quad \quad \quad CP-SSL-MOS&  Wav2Vec2.0 Base&\twemoji{multiply}&0.370&0.668&0.635&0.475&0.036&\textbf{0.952}&0.939&0.813  \\
  
 \quad \quad \quad \quad PH-SSL-MOS&  Wav2Vec2.0 Base&\twemoji{multiply}&0.344&0.669&0.629&0.474&0.241&0.951&0.939&0.806 \\
 
 \quad \quad \quad \quad PH-SSL-MOS&  Wav2Vec2.0 Large&\twemoji{multiply}&0.320&0.689&0.651&0.489&0.035&0.935&0.933&0.787 \\

\quad \quad \quad \quad PH-SSL-MOS&  HuBERT Base&\twemoji{multiply}&0.338&0.685&0.653&0.489&0.051&0.949&0.935&0.806 \\
 
  \quad \quad \quad \quad S-SSL-MOS& HuBERT Base&\twemoji{multiply}&0.966&0.656&0.627&0.465&0.620&0.945&0.931&0.798 \\

  \hline  
  
  \textbf{PS-SQA}& \- & \twemoji{check mark} &\textbf{0.320}&0.685&\textbf{0.672}&\textbf{0.506}&\textbf{0.027}&0.944&\textbf{0.953}&\textbf{0.840}\\

  \quad \quad \quad \quad CP-SSL-MOS&  Wav2Vec2.0 Base&\twemoji{check mark}&0.492&0.679&0.641&0.480&0.143&0.954&0.944&0.828  \\
  
 \quad \quad \quad \quad PH-SSL-MOS&  Wav2Vec2.0 Base&\twemoji{check mark}&0.374&0.654&0.627&0.471&0.040&\textbf{0.959}&0.945&0.828 \\
 
 \quad \quad \quad \quad PH-SSL-MOS&  Wav2Vec2.0 Large&\twemoji{check mark}&0.328&0.686&0.651&0.489&0.042&0.941&0.938&0.802\\

\quad \quad \quad \quad PH-SSL-MOS&  HuBERT Base&\twemoji{check mark}&0.341&\textbf{0.687}&0.654&0.489&0.054&0.950&0.938&0.813 \\
 
  \quad \quad \quad \quad S-SSL-MOS& HuBERT Base&\twemoji{check mark}&0.477&0.660&0.623&0.465&0.159&0.908&0.931&0.793 \\

  \hline

 \end{tabular}
\end{table*}

\begin{table*}[t]
 \setlength{\belowcaptionskip}{-0.4cm}
 \caption{Experimental results of the official baseline, the participating teams, and our proposed PS-SQA on the test set of SingMOS. The \textbf{bold} and \underline{underline} numbers indicate the optimal and sub-optimal results, respectively. 
 }
\label{table:ensemble}
 \centering
 \begin{tabular}{c c c c c |c c c c } 
  \hline
  \multirow{2}{*}{\textbf{Participating Systems}}&\multicolumn{4}{c|}{\textbf{Utterance-level Metrics}} &\multicolumn{4}{c}{\textbf{System-level Metrics}} \\
  \cline{2-9}
   & MSE $\downarrow$ & LCC $\uparrow$ & SRCC $\uparrow$ & KTAU $\uparrow$ & MSE $\downarrow$ & LCC $\uparrow$ & SRCC $\uparrow$ & KTAU $\uparrow$
   \\
   
\hline 
    B01 (Offical Baseline)& 0.419&0.594&0.605&0.442&0.079&0.851&\underline{0.859}&\underline{0.687}\\
T01&0.366	&0.605&	0.603&	0.440&	\underline{0.051}&	0.858&	0.837&	0.684
\\
T03& 0.432&	0.597&	0.583&	0.426	&0.061&	0.848&	0.819&	0.637
\\
T04& \underline{0.363}&	0.624&	0.604&	0.445&	0.056&	\underline{0.869}&	0.833&	0.640
\\
T05& \underline{0.363}&0.609&	0.593&	0.434&	0.069&	0.791	&0.807	&0.657
\\
T06& \textbf{0.358}&\textbf{0.637}&\underline{0.625}&\underline{0.460}&	0.063&	0.841&	0.831	&0.657
\\

   T08 (Our Submission)& 0.384&0.628&0.620&0.455&0.072&0.845&0.856&\underline{0.687}\\
   \textbf{PS-SQA}& 0.452&\underline{0.634}&\textbf{0.639}&\textbf{0.470}&\textbf{0.031}&\textbf{0.880}&\textbf{0.888}&\textbf{0.717}\\

  \hline
 \end{tabular}
\end{table*}

\subsection{Comparsion among Different SSL-based MOS Predictors}
\label{Comparsion among Different SSL-based MOS Predictors} 

First, we compared the performance of the proposed pitch-aware and spectrum-aware SSL-based MOS predictors as well as the plain SSL-based MOS predictor to validate the effectiveness of the introduced pitch and spectral information for singing quality assessment. 
We selected 5 pre-trained models to use in these predictors, including Wav2Vec2.0 Base, Wav2Vec2.0 Large, HuBERT Base, HuBERT Large and HuBERT Extra Large. 
Therefore, a total of 20 predictors were compared, as shown in Table \ref{table:main experiment}. 
For the spectrum-aware SSL-based MOS predictor, we adopted a no-quantized APCodec pre-trained on the VCTK-0.92 corpus \cite{yamagishi2019cstr} and a 2-layer Conformer, and the output feature dimension of spectral-level acoustic features was 64. 
During training, we trained all predictors for 1,000 epochs, with a batch size of 4 and optimizer as stochastic gradient descent (SGD) with a learning rate of 0.0001. 
For the checkpoint saving strategy, we followed the same approach as UTMOS \cite{saeki2022utmos}, selecting the best system-level SRCC checkpoint calculated from the validation set. 
If the system-level SRCC didn’t decrease within 15 epochs, early stopping was applied. 
The purpose of comparing these predictors, besides validating the effectiveness of the introduced pitch and spectral information, is to select models for fusion. 
Hence, the experiments were conducted on the validation set. 

The experimental results are shown in Table \ref{table:main experiment}. 
For the two pitch-aware SSL-based MOS predictors, the pitch-histogram-aware SSL-based MOS predictor (i.e., PH-SSL-MOS in Table \ref{table:main experiment}) demonstrated more stable performance. 
Compared to the plain SSL-based MOS predictor (i.e., SSL-MOS in Table \ref{table:main experiment}), the pitch-histogram-aware one showed significant improvements in most metrics regardless of the SSL model used. 
Unfortunately, the compressed-pitch-aware SSL-based MOS predictor (i.e., CP-SSL-MOS in Table \ref{table:main experiment}) seems to be sensitive to the SSL model. 
Specifically, when using Wav2Vec2.0 Large, its performance at both the utterance and system levels is very poor. 
The experimental results above confirm the effectiveness of introducing pitch-related information for singing quality assessment. 
Among the methods, the pitch histogram is more suitable for this task compared to pitch values. 
This may be attributed to the pitch histogram's better representation of the characteristics of singing voices, confirming the hypothesis in Section \ref{pitch-histogram}. 
For the spectrum-aware SSL-based MOS predictor (i.e., S-SSL-MOS in Table \ref{table:main experiment}), when using Wav2Vec2.0 Large, its performance is also poor, and regardless of the SSL model used, its MSE metric significantly increases compared to the plain predictor. 
Nevertheless, its performance is satisfactory regarding ranking metrics, such as system-level SRCC. 
Therefore, the introduction of spectral-level acoustic information is helpful in improving the accuracy of singing quality assessment models to a certain extent.

\subsection{Validation on Model Fusion and Bias Correction}
\label{Selection of Fusion Models and Validation of Bias Correction}
\vspace{-1.5mm}
As mentioned in Section \ref{fusion}, the PS-SQA employed a model fusion strategy for better singing MOS prediction accuracy. 
We selected five predictors from the twenty predictors shown in Table \ref{table:main experiment} for fusion and constructed the PS-SQA. 
Since system-level SRCC is an important ranking metric in VoiceMOS Challenge 2024, we selected the top five predictors based on their system-level SRCC rankings. 
The selected predictors are listed in Table \ref{table:sub experiment}, including 1) compressed-pitch-aware MOS predictor with Wav2Vec2.0 Base as the SSL model, 2) pitch-histogram-aware MOS predictor with Wav2Vec2.0 Base as the SSL model, 3) pitch-histogram-aware MOS predictor with Wav2Vec2.0 Large as the SSL model, 4) pitch-histogram-aware MOS predictor with HuBERT Base as the SSL model, and 5) spectrum-aware MOS predictor with HuBERT Base as the SSL model. 
We can observe that among the five selected predictors, four incorporate pitch information, while only one incorporates spectrum information.
This further indicates that pitch information is more important for enhancing the performance of singing quality assessment methods compared to spectrum information. 
This may be due to the specific characteristics of singing voices. 
Within the pitch-aware methods, three predictors based on pitch histograms were selected, while only one predictor based on compressed pitch was chosen. 
This indicates that pitch histograms can provide more easily interpretable pitch information for SSL-based MOS prediction models, facilitating the learning process for the models. 
As the results of PS-SQA without bias correction in Table \ref{table:sub experiment} (i.e., the rows annotated with \ding{53} in the ``Bias Correction" column) suggested, except for the system-level LCC metric, the performance of the fused PS-SQA is higher than any of the individual predictors in all other metrics. 

\begin{figure}
 \setlength{\belowcaptionskip}{-0.8cm}
\begin{center}
\subfigure[]{\includegraphics[width=0.49\linewidth]{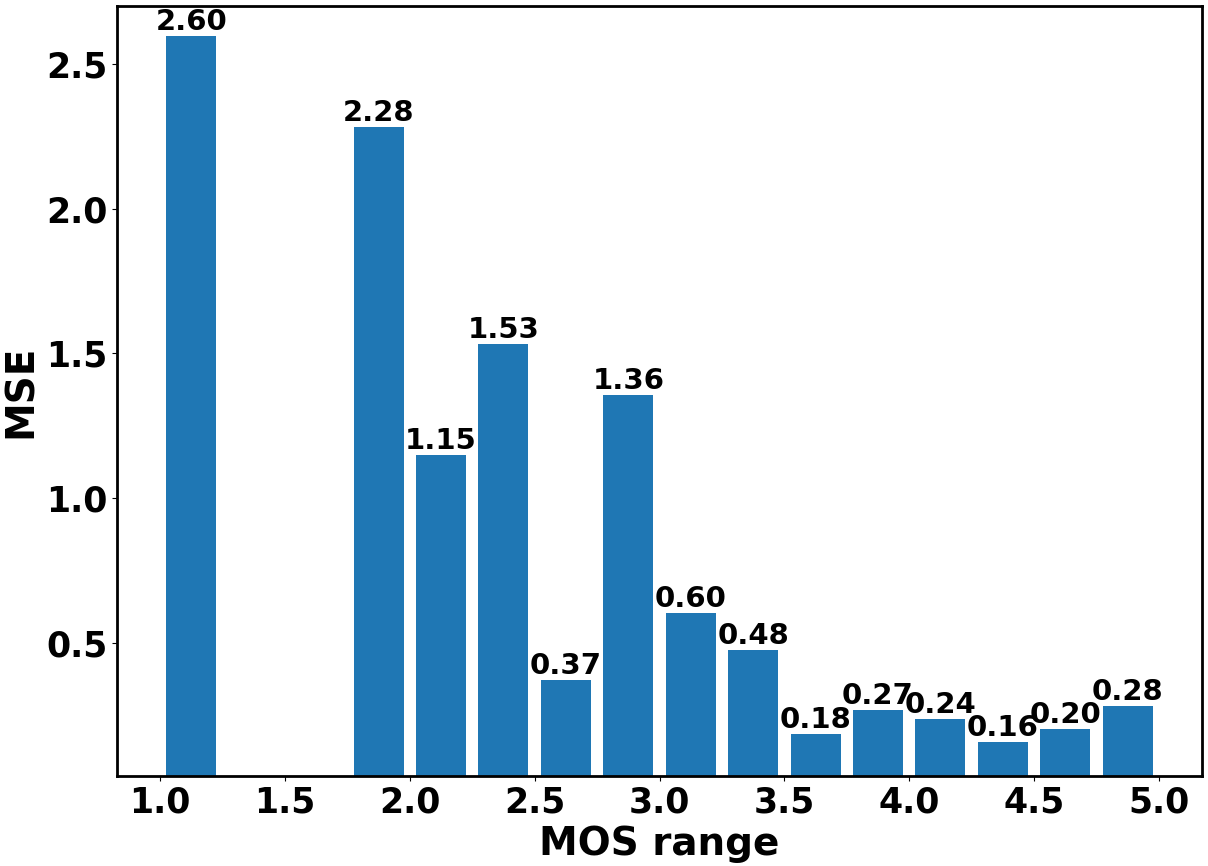}}
\subfigure[]{\includegraphics[width=0.49\linewidth]{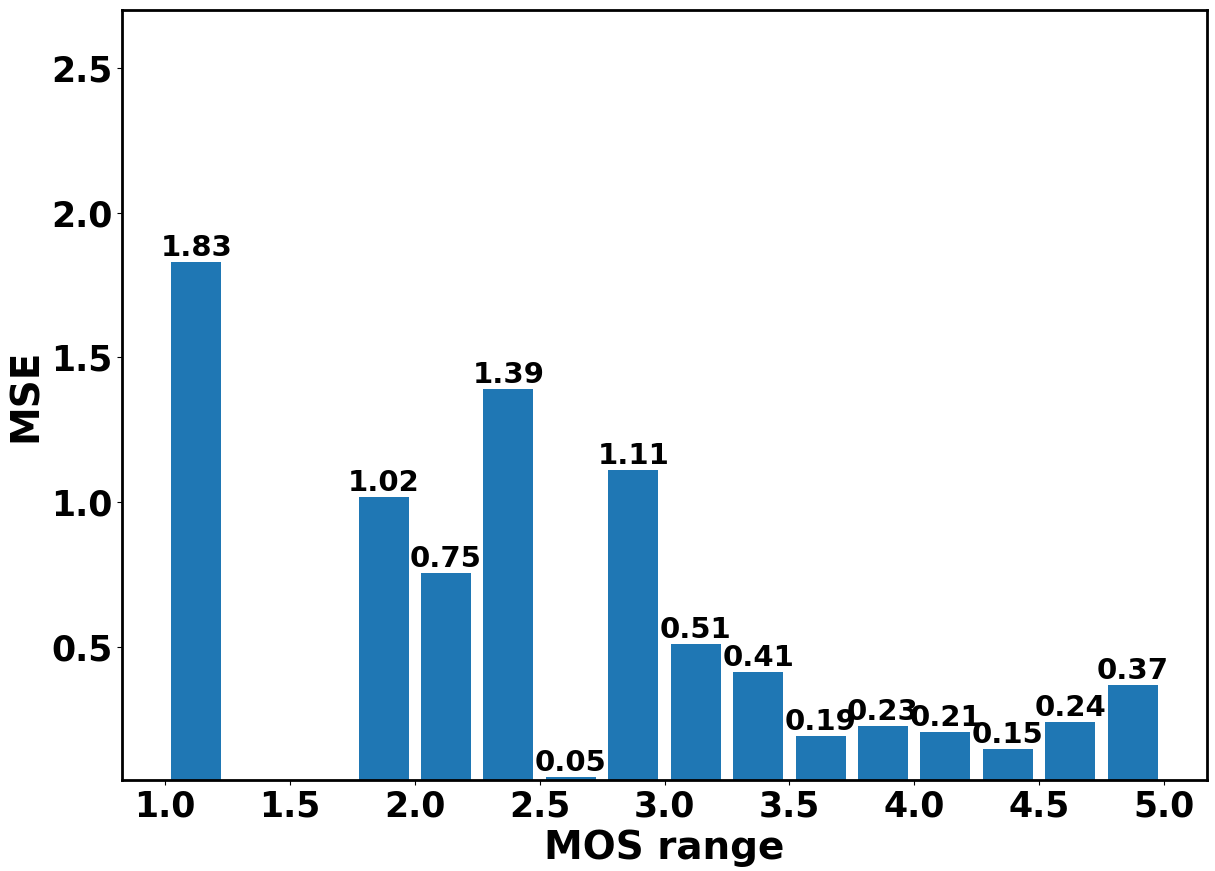}}
\caption{Bar charts showing the MSE for each MOS score segment (interval length of 0.25) for pitch-histogram-aware MOS predictor with Wav2Vec2.0 Base as the SSL model, where subplot (a) displays the results without bias correction and subplot (b) displays the results with bias correction.}
\label{distribute of dataset2}
\end{center}
\end{figure}

Specifically, after model fusion, the system-level SRCC improved by 0.013 compared to the best individual predictor.
This indicates that the model fusion strategy combines the strengths of individual MOS predictors, achieving more accurate singing MOS predictions. 
This confirms the effectiveness of model fusion.

Then, we added the bias correction branch to each of the individual predictors to explore the role of bias correction. 
The experimental results are also listed in Table \ref{table:sub experiment}. 
By comparing the performance of individual predictors with (i.e., the rows annotated with \twemoji{check mark} in the ``Bias Correction" column) and without the bias correction branch, it can be observed that the introduction of the bias correction branch effectively improved performance in most metrics. 
Figure \ref{distribute of dataset2} also shows the MSE metrics for pitch-histogram-aware MOS predictor with Wav2Vec2.0 Base as the SSL model across different MOS segments using bar charts, comparing the results without and with bias correction. 
The label MOS scores in the validation set, ranging from 1 to 5, are evenly divided into 16 segments with an interval of 0.25. 
There are no samples in the 2nd and 3rd MOS segments.
By comparing the two subplots in Figure \ref{distribute of dataset}, it is clear that in the low MOS segments, the introduction of bias correction significantly reduced the MSE metric. 
In the MOS segment from 1.75 to 2, the MSE reduction even exceeded 1. 
This indicates that bias correction effectively enhances the predictor's modeling capability on low-resource data, effectively mitigating the difficulty the predictor faces when handling specific MOS segments. 
When these five bias-corrected predictors are fused to construct the PS-SQA, the performance also surpassed that of the individual predictors. 
The bias-corrected fused PS-SQA also shows a slight improvement in system-level metrics compared to the non-bias-corrected fused PS-SQA.

\vspace{-2mm}
\subsection{Position in VoiceMOS Challenge 2024 Competition Systems} 

Based on the results discussed in Sections \ref{Comparsion among Different SSL-based MOS Predictors} and \ref{Selection of Fusion Models and Validation of Bias Correction} on the validation set, we chose the PS-SQA with bias correction and model fusion strategies as our competitive system to compare with the systems participating in track 2 of VoiceMOS Challenge 2024. 
This allows us to evaluate the position of our proposed method among these competition systems. 
Table \ref{table:ensemble} presents the experimental results of the official baseline, the participating teams, and the PS-SQA proposed in this paper on the SingMOS evaluation set. 
Our submitted system (i.e., T08) significantly outperformed the other participating systems (i.e., T01, T03, T04, T05 and T06) in terms of the system-level SRCC metric used for ranking, although it was slightly inferior to the official baseline. 
Our submitted system only utilized the pitch-histogram-aware SSL-based MOS predictor as shown in Figure \ref{PA/SA-SSL-MOS}(b) without layer normalization. 
Building on this, we made improvements and proposed other-information-aware SSL-based MOS predictors. 
We also introduced the bias correction branch in various predictors and adopted a model fusion strategy. 
Currently, in terms of system-level metrics, the proposed PS-SQA significantly outperformed all other systems. 
Specifically, compared to our submitted system, PS-SQA improved the system-level SRCC by 0.032.

\vspace{-2mm}
\section{Conclusion}
\label{sec: Conclusion}

This paper proposes a novel pitch-and-spectrum-aware singing quality assessment method, called PS-SQA, which is an improvement version of the system we submitted to track 2 of VoiceMOS Challenge 2024. 
The PS-SQA first introduces multiple MOS predictors that incorporate pitch and spectrum-related information into the SSL-based MOS prediction model, tailored to the characteristics of singing voices. 
The PS-SQA then selects these predictors for model fusion and introduces a bias correction branch to overcome training bias caused by low-resource training data. 
Experimental results confirm that our proposed PS-SQA significantly outperforms all participating systems in terms of system-level evaluation metrics. 
Enhancing the MOS prediction accuracy of PS-SQA by introducing more types of predictors will be our future work.

\bibliographystyle{IEEEbib}
\bibliography{strings,refs}

\end{document}